\documentclass[10pt,a4paper]{article}
\newcommand{\be}{\begin{equation}}
\newcommand{\ee}{\end{equation}}
\newcommand{\nbar}[1]{\overline{#1}}  
\newcommand{\xprim}{\acute{x}}
  
\newcommand{\cq}{{\cal Q}}

\def\bea{\begin{eqnarray}}
\def\eea{\end{eqnarray}}
\def\beas{\begin{eqnarray*}}
\def\eeas{\end{eqnarray*}}

\def\parp{{\partial}^{+}}
\usepackage{latexsym}

\usepackage{amscd}
\usepackage{pb-diagram}
\usepackage{psfrag}
\usepackage{graphicx}
\usepackage[latin1]{inputenc}
\usepackage{amsfonts}
\usepackage{cite}
\usepackage{stmaryrd}
\usepackage{amsmath}
\usepackage{fancyhdr}
\usepackage{setspace}
\usepackage{epic}
\usepackage{lscape,array}
\usepackage{amsthm,latexsym,multibox}
\usepackage{amssymb,amsfonts,amsbsy}

\input epsf

\begin{document}

\begin{centering}

  \vspace*{-2cm}

  \textbf{\Large{Particle Physics as Representations of the \\
Poincar\'{e} Algebra}}

  \vspace{2.5cm}

  {\large Lars Brink$^{a}$ }
  \vspace{1cm}

  \begin{minipage}{.8\textwidth}\small

\mbox{}\kern-4pt$^a$Department of Fundamental Physics, Chalmers University of Technology, S-412 96 G\"oteborg, Sweden, tfelb@fy.chalmers.se

\end{minipage}

\end{centering}

\vspace{2cm}

\begin{center}
  \begin{minipage}{.9\textwidth}
    \textsc{Abstract}. Eugene Wigner showed already in 1939 that the elementary particles are related to the irreducible representations of the Poincar\'e algebra. In the light-cone frame formulation of quantum field theory one can extend these representations to depend also on a coupling constant. The representations then become non-linear and contain the interaction terms which are shown to have strong uniqueness. Extending the algebra to supersymmetry it is shown that two field theories stick out, $N=4$ Yang-Mills and $N=8$ Supergravity and their higher dimensional analogues. I also discuss string theory from this starting point.

  \end{minipage}
\end{center}

\vspace{3cm}

{\it Lecture presented at the
  Poincar\'e Symposium held in Brussels
  on October 8-9,  2004}

\vfill

\noindent
\mbox{}
\raisebox{-3\baselineskip}{%
  \parbox{\textwidth}{\mbox{}\hrulefill\\[-4pt]}}
{\scriptsize}

\thispagestyle{empty}

\newpage

\pagestyle{myheadings}
\markboth{\textsc{\small authors}}{%
  \textsc{\small Light-Cone Field Theory}}
\addtolength{\headsep}{4pt}

\section{Introduction}
It  took the genius of Henri Poincar\'e to realize in 1905 that the symmetry underlying special relativity formed a group~\cite{Poincare} and hence was amenable to a formidable machinery of mathematics. Then in his famous paper in 1939 Eugene Wigner~\cite{Wigner:1939cj} showed that the irreducible representations of the Poincar\'e algebra can be classified by the spin of the representations (as well as the mass). The spin can be either integer or half-integer and we find naturally the bosons and the fermions on an equal footing. A quantum field theory must be invariant under this symmetry, and that is certainly one of the starting points for building a quantum field theory. Another one is, of course,  gauge invariance, which introduces unphysical degrees of freedom and shadows the consequences of the Poincar\'e symmetry. There are, however, gauge choices for which the unphysical degrees of freedom can be integrated out of the functional integral and then the remaining symmetry is just the Poincar\'e one. These gauge choices are typically such that the light-cone components of the gauge field be zero. By also interpreting one of the light-cone directions as the evolution parameter, the "time",  one finds that the unphysical degrees of freedom satisfy algebraic equations and hence can be integrated out from the functional integral~\cite{Brink:1982pd}. This process will introduce new interaction terms and, as we will see, also some mild non-locality into the interacting Lagrangians. The Poincar\'e invariance is now obscured since some of the covariance is lost and it will indeed be non-linearly implemented. The generators can be found by introducing the solutions of the unphysical degrees of freedom into them and by making a gauge transformation to make sure one stays in the gauge.
  
 There is an obvious alternative way to construct these gauge fixed Lagrangians, namely to take Wigner's approach very literally. Poincar\'e invariance is a physical symmetry, a global symmetry that we can test,  and which we believe must be an underlying symmetry of any theory. In some sense the gauge invariances are artificial. It is really only their global limits that we measure physically. Hence we only need to implement the global symmetries. If we construct a free quantum field which describes the physical degrees of freedom consistent with the Poincar\'e invariance and other global symmetries, we can try to construct interaction terms with a coupling constant to extend the generators such that they close again~\cite{Bengtsson:1983pd}. This means that these representations can be characterized by a parameter, the coupling constant. From the arguments above it is clear that the process must work at least for the known quantum field theories, and hence could be used to find new theories as well as to argue about the uniqueness of the known ones.
 
In this talk I will show that all massless theories can be found in the latter way and that the procedure is very general. Among the theories discussed, we see that the spin-$2$ theory is just another field theory. There will be no hint of the equivalence principle or the covariance principle and the symmetry will be strictly just the Poincar\'e one. Since this formulation will contain infinitely many interaction terms which have to be found by the procedure, it is obvious that only the first few terms can be found. The formulation will be useless for most purposes except questions about simple loop graphs in the quantum theory. Would this procedure have produced the Einstein theory on a different planet? This we can only speculate about, but it is possible. The process is more important when we turn to supersymmetric theories. Here it is seen that the $N=4$ Yang-Mills and $N=8$ Supergravity theories stick out as very special theories indicating that there must be a close relationship among them. I will also show that the formalism is very natural to construct the corresponding theories in $d=10$ and $d=11$~\cite{Ananth:2004es},~\cite{Ananth:2005vg}.

Finally one can also use this procedure to extend the representations to be written in terms of functionals, In this way one can show that there exist indeed Poincar\'e invariant string theories but with very little freedom to construct such ones.
 
\section{Light-Frame Formulation of  Field Theories}
 
 We know since the time of Poincar\'e, Lorentz and Einstein that relativistic dynamics is invariant under the Poincar\'e algebra. We start by working in a four-dimensional space with translations $P^\mu$ and rotations and boosts $J^{\mu\nu}$, where $\mu=0, 1,2,3$ and the metric is  $\eta_{\mu\nu}=(-1,1,1,1)$. (We will use both capital and lower case letters for the generators. The difference will be clear later.) The algebra is 
 
 \bea  
[P^\mu , P^\nu]	& = & 0 \ , \\  \cr
[ J^{\mu \nu} , P^\sigma ] & =	& i (\eta^{\mu \sigma} P^\nu -\eta^{\nu\sigma} P^\mu)\ , \\  \cr
[ J^{\mu \nu} , J^{\alpha \beta} ] & = & i (\eta^{\mu \alpha} J^{\nu\beta}
 + \eta^{\alpha \nu} J^{\beta \mu} + \eta^{\nu \beta} J^{\mu \alpha}
+ \eta^{\beta \mu} J^{\alpha \nu}) .
\eea
  
In this talk I will study representations of this algebra and show that we can find all relativistic field theories in a systematic way in this study.  
 
 In his famous paper of 1949 Paul Dirac~\cite{Dirac:1949cp} argued that  for a relativitistically invariant theory  any direction within the light-cone can be the evolution parameter, the "time". In particular we can use one of the light-cone directions. For this discussion we will use ${x^{+}}=\frac{1}{\sqrt 2}\,(\,{x^0}\,{+}\,{x^3})$ as the time. The coordinates and the derivatives that we will use will then be
 
 \bea
{x^{\pm}}&=&\frac{1}{\sqrt 2}\,(\,{x^0}\,{\pm}\,{x^3}\,)\ ;\qquad ~ {\partial^{\pm}}=\frac{1}{\sqrt 2}\,(\,-\,{\partial_0}\,{\pm}\,{\partial_3}\,)\ ; \\
x &=&\frac{1}{\sqrt 2}\,(\,{x_1}\,+\,i\,{x_2}\,)\ ;\qquad  {\bar\partial} =\frac{1}{\sqrt 2}\,(\,{\partial_1}\,-\,i\,{\partial_2}\,)\ ; \\
{\bar x}& =&\frac{1}{\sqrt 2}\,(\,{x_1}\,-\,i\,{x_2}\,)\ ;\qquad  {\partial} =\frac{1}{\sqrt 2}\,(\,{\partial_1}\,+\,i\,{\partial_2}\,)\ ,
\eea
so that 
\be
{\parp}\,{x^-}={\partial^-}\,{x^+}\,=\,-\,1\ ;\qquad {\bar \partial}\,x\,=\,{\partial}\,{\bar x}\,=+1 \ .
\ee
 
The derivatives are, of course, related to the momenta through the usual formula $p^{\mu}= -i \partial^{\mu}$ and we use the light-cone decomposition also for $p^{\mu}$. We will only consider massless theories so we solve the condition $p^2=0$. We then find 
\be 
p^- = \frac{p\bar p}{p^+}.
\ee
 
The generator $p^-$ is really the Hamiltonian conjugated to the light cone time $x^+$ and we see that the translation generators of the Poincar\*e algebra are written with just three operators. We will use Dirac's vocabulary that generators that involve the "time" are called dynamical (or Hamiltonians) and the others kinematical. Using light-cone notation and the complex one from above for the transverse directions,  the most general form of the generators of the full Poincar\'e algebra at $x^+=0$ is then given by the four momenta

\be
p^-_{}~=~-i\frac{\partial\bar\partial}{\partial^+_{}}\ ,\qquad p^+_{}~=~-i\,\partial^+_{}\ ,\qquad p~=~-i\,\partial\ ,\qquad \bar p~=~-i\,\bar\partial\ ,
\ee
the kinematical transverse space rotation 

\be
j~=j^{12}=~x\,\bar\partial-\bar x\,\partial + \lambda,
\ee
the other kinematical generators 

\be
j^+_{}~=~i\, x\,\partial^+_{}\ ,\qquad \bar j^+_{}~=~i\,\bar x\,\partial^+_{}\ ,
\ee
and 
\be
 j^{+-}_{}~=~i\,x^-_{}\,\partial^+,
\ee
as well as the dynamical boosts 

\bea
j^-_{}&=&i\,x\,\frac{\partial\bar\partial}{\partial^+_{}} ~-~i\,x^-_{}\,\partial_{}\ + i \lambda \frac{\partial}{\partial^+} ,\qquad \\
\bar j^-_{}&=&i\,\bar x\,\frac{\partial\bar\partial}{\partial^+_{}}~ -~i\,x^-_{}\,\bar\partial_{}\ + i \lambda \frac{\bar \partial}{\partial^+}.
\eea

There is  one degree of freedom in the algebra, namely the parameter $\lambda$ which is the helicity. At this stage it is arbitrary and checking the corresponding spin one finds, of course, that it is $|\lambda|$. Hence the algebra covers all possible free field theories. We can let the generators act on a complex field $\phi(x)$ with helicity $\lambda$, with its complex conjugate having the opposite helicity. This is the "first-quantized" version. We can also consider the fields as operators having  the commutation relation.

\be
[ \partial^+ \bar\phi(x), \phi(x')] = -\frac{i}{2} \delta(x-x'),
\ee
 where hence the momentum field conjugate to $\phi$ is $\partial^+ \bar \phi$.
 
We then introduce the "second-quantized" representation O in terms of the  "first-quantized" representation o as $O = 2i \int d^4x \partial^+ \bar\phi(x) ~o~ \phi(x)$. We then find that the commutator between two of the generators $J_1$ and $J_2$ is
 
\be
[J_1,J_2] = 2 i \int d^4x \partial^+ \bar\phi(x) [j_1,j_2] \phi(x).
\ee

We can  understand that $P^-$ truly is the Hamiltonian using equ.(8)

\be
P^- = 2 \int d^4x \partial^+\bar\phi(x) \frac{\partial\bar \partial}{\partial^+}\phi(x).
\ee

Legendre transforming to the Lagrangian using the field momenta  from equ.(15) we get the action

\bea
S &=&  \int d^4x[\partial^+\bar\phi(x) \partial^- \phi(x) +  \partial^+\phi(x) \partial^- \bar\phi(x) - 2 \partial^+\bar\phi(x) \frac{\partial \bar \partial}{\partial^+}\phi(x)] \nonumber \\
 &=& \int d^4x\partial^+\bar\phi(x) \Box \phi(x) .
\eea
It is remarkable that there is a unique form of the kinematic term for any spin-$\lambda$ field. We should remember though that to specify the theory we have to give all Poincar\'e generators, since the action via the Hamiltonian is just one of those generators. They will show what spin the field describes.

In this representation it is straightforward to try to add interaction terms to the Hamiltonian. This was done in \cite{Bengtsson:1983pd}. Every dynamical generator will have interaction terms. The procedure is very painstaking  and there are as far as I know no other way than trial and error to find the non-linear representation. On the other hand, once such a representation is found it represents a possible relativistically invariant interacting field theory. The result is that for every integer $\lambda$ there exists  a possible three-point interaction. For $\lambda$ even, the unique solutions are

\bea S &= &\int d^4x\Bigl\{\bar\phi(x) \Box \phi(x)  \nonumber\\
& &+ g \Bigl[ \sum_{n=0}^\lambda (-1)^n {\lambda \choose n} \bar\phi(x) {\partial^+}^\lambda (\frac{{\bar \partial}^{\lambda - n}}{{\partial^+}^{\lambda - n} } \phi(x)    \frac{{\bar \partial}^{\lambda}}{{\partial^+}^{\lambda } } \phi(x) ) + c.c.\Bigr] \Bigr\} \nonumber\\ &  &+O(g^2).
\eea

For $\lambda$ odd, the field $\phi(x)$ must be in the adjoint representation of an external group $\phi^a(x)$ and we have to introduce the fully antisymmetric structure constants $f^{abc}$ in the interaction terms to find a possible term. The results is 

\bea S &= &\int d^4x\Bigl\{\bar\phi^a(x) \Box \phi^a(x)\nonumber \\
& &+ gf^{abc} \Bigl[ \sum_{n=0}^\lambda (-1)^n {\lambda \choose n} \bar\phi^a(x) {\partial^+}^\lambda (\frac{{\bar \partial}^{\lambda - n}}{{\partial^+}^{\lambda - n} } \phi^b(x)   \frac{{\bar \partial}^{\lambda}}{{\partial^+}^{\lambda } } \phi^c(x) ) + c.c.\Bigr ] \Bigr\} \nonumber\\&  &+O(g^2).
\eea
We note the non-locality in the interaction term in terms of inverses of $\partial^+$. The easiest way to understand it is to Fourier transform to momentum space. In the calculations it is really defined by the rule $\frac{1}{\partial^+} \partial^+ f(x^+) = f(x^+)$. When performing a calculation one has to specify exactly the situation of the pole in  $\partial^+$. In an sense this is a remainder of the gauge invariance.

We can now check for special values of $\lambda$.

\vspace{3mm}

$\bullet \hspace{3mm}{\it \lambda =0}$

\vspace{3mm}

The dimension of the coupling constant $g$ is $1$ (in mass units) and this is the usual $\phi^3$- theory. This theory is superrenormalizable but not physical since it does not have a stable vacuum having a potential with no minimum.
\vspace{3mm}

$\bullet \hspace{3mm}{\it \lambda =1}$

\vspace{3mm}

The dimension of the coupling constant $g$ is $0$ and this theory is nothing but non-abelian gauge theory in a specific gauge. If we go on we know that we need a four-point coupling to fully close the algebra. Note that the action has no local symmetry and the gauge group only appears as the external symmetry group.
\vspace{3mm}

$\bullet \hspace{3mm}{\it \lambda =2}$

\vspace{3mm}

The dimension of the coupling constant $g$ is $-1$ and this theory is the beginning series of a gravity theory. It is clear from the dimensions of the coupling constant that interaction terms to arbitrary order can be constructed without serious non-localities. The four-point function related to Einstein's theory is known~\cite{Bengtsson:1983vn}. Going beyond the four-point coupling is probably too difficult, unless powerful computer methods could be devised. We expect several solutions, of course, since we know that the Hilbert action is but the simplest of all actions consistent with the equivalence principle. Note that the action above, which is a fully gauge fixed Hilbert action expanded in the fluctuations around the Minkowski metric, has no local symmetry, no covariance and knows nothing about curved spaces. It is probably useless for discussions about global properties of space and time but can be useful in the study of quantum corrections; to understand the finiteness properties of the quantum theory.
\vspace{3mm}

$\bullet \hspace{3mm}{\it \lambda > 2}$

\vspace{3mm}

The dimension of the coupling constant $g$ is $< -1$ and these theories are theories for higher spins. Again they are non-renormalizable in the naive sense like the the spin-$2$ theory above. There are strong reason to believe that these theories cannot be Poincar\'e invariant one by one when we go to higher orders in the coupling constant, but the result above is an indication that certain sums of such theories interacting with each other could possibly be invariant theories.

We can also find interacting solutions for $\lambda$ half-integer. We can, of course, not have a three-point coupling. We will in fact not be able to find self-interacting theories but have to consider the coupling of the half-integer spin field to an integer spin field. We then find that we can couple a spin-$\frac{1}{2}$ field to a spin-$1$ or a spin-$0$ field to recover in the first case a non-abelian gauge field coupled to a spin-$\frac{1}{2}$ field $\psi^i(x)$ in a representation characterized by $i$ of the external group such that we can have a coupling $\bar \psi_i \psi^j \phi^a C^i_{ja}$, with $ C^i_{ja}$ the Clebsch-Gordan coefficient. It is interesting to note that it is only in the interacting theory that we can prove the spin-statistics theorem~\cite{Bengtsson:1984gu}. The formalism demands the spin-$\frac{1}{2}$ field to be of odd Grassmann type and the integer spin fields to be even. Note that there is no spinor space. The spin-$\frac{1}{2}$ field is a complex (Grassmann odd) field with no space-time index. Its equation of motion looks just like the one for a bosonic field. (Remember the free equation the follows from equ. (18).) However, the dimension of the field $\psi(x)$ is different from the one of the bosonic field, so the free action is
\be
S= \int d^4x\partial^+\bar\psi(x) \frac{\Box}{\partial^+} \psi(x) .
\ee
The fact that we do not need to use spinors is very special for $d=4$, since the transverse symmetry which is covariantly realized is $SO(2)\approx U(1)$, which does not distinguish spinor representations.

We have hence seen that we can find all known unitary relativistic field theories as representations of the Poincar\'e algebra, and we see their uniqueness and also what kind of possibilities there are for higher spin fields.  In a gauge invariant formulation one can attempt to add in new terms that are gauge invariant. Invariably they lead to problems with unitarity. We do not see those terms here since the theories are unitary by construction.

\section{Light-Frame Formulation of  Supersymmetric Field Theories}

I think that Poincar\'e would have been very interested in supersymmetry. It is an extension of the Poincar\'e algebra and hence a restriction on relativistic dynamics. It is true that the world does not look supersymmetric as such, but a good working hypothesis is that at some stage supersymmetry is indeed a symmetry of the world.

Supersymmetry is an augmentation of the Poincar\'e algebra with a spinor generator $Q_\alpha$ with the anti-commutator 

\be
\{ Q_\alpha, \bar Q_\beta \} = \gamma^{\mu}_{\alpha \beta} P_\mu.
\ee
The spinor $Q_\alpha$ is  four-component. It satisfies the so-called Majorana condition which makes it real in a certain representation of the $\gamma$-matrices.  In the light-cone frame the spinor splits up into two two-component spinor that can be rewritten as two complex operators, which we call $Q_+= -\frac{1}{2} \gamma_+ \gamma_- Q$ and $Q_- =- \frac{1}{2} \gamma_- \gamma_+ Q$. From the Clifford algebra $\{\gamma^\mu, \gamma^\nu\}= 2 \eta^{\mu \nu}$ with $\eta = diag(-1,1,1,1)$ we see that $Q=Q_+ + Q_-$, and that the products  $-\frac{1}{2} \gamma_+ \gamma_- $ and $-\frac{1}{2} \gamma_- \gamma_+$ are projection operators. We can linearly combine the two components of the spinors into complex entities with no indices.
We can also augment by letting the $Q$'s transform as the representation $\bf N$ under $SU(N)$. The light-cone supersymmetry algebra is then

\bea
\{Q_+^m,{ \bar Q}_{+n}\}& =& -\sqrt {2}  \delta^m_n P^+ \\
\{Q_-^m,{ \bar Q}_{-n}\}& =& -\sqrt {2}  \delta^m_n P^- \\
\{Q_+^m,{ \bar Q}_{-n}\} &=& -\sqrt {2}  \delta^m_n P,
\eea
where all other anticommutators are zero, except for the complex conjugate of the last one. The indices $m,n$ run from $1$ to $N$.

The superPoincar\'e algebra can now be represented on a superspace with coordinates $ x^{\pm}, x, \bar x, \theta^m, \bar \theta_n$, where the coordinates $\theta^m$ and $ \bar \theta_n$ are complex conjugates, Grassmann odd and transform as  $\bf N$ and $\bf \bar N$ under $SU(N)$. We will denote their derivatives as
\be
{{\bar \partial}_m}\,~\equiv~\,\frac{\partial}{\partial\,{\theta^m}}\ ;\qquad{\partial^m}\,~\equiv~\,\frac{\partial}{\partial\,{\bar \theta}_m}\ .
\ee

The $Q$'s are then represented as (We use the notation with lower case letters for operators that act on the field.)

\bea
{q_+^m}=-{\partial^m}\,+\,\frac{i}{\sqrt 2}\,{\theta^m}\,{\partial^+}\ ;\qquad{{ \bar q}_{+n}}=\;\;\;{{\bar \partial}_n}\,-\,\frac{i}{\sqrt 2}\,{{\bar \theta}_n}\,{\partial^+}\ ,
\eea

and the dynamical ones as

\be
{q}^m_{\,-}~=~\frac{\bar \partial}{\partial^+_{}}\, q^{\,m}_{\,+}\ ,\qquad 
{\bar{q}}_{\,-\,m}^{}~=~\frac{\partial}{\partial^+_{}}\, \bar q_{\,+\,m}^{}\ .
 \ee

On this space we can also represent "chiral" derivatives anticommuting with the supercharges $Q$.

\bea
{d^{\,m}}=-{\partial^m}\,-\,\frac{i}{\sqrt 2}\,{\theta^m}\,{\partial^+}\ ;\qquad{{\bar d}_{\,n}}=\;\;\;{{\bar \partial}_n}\,+\,\frac{i}{\sqrt 2}\,{{\bar \theta}_n}\,{\partial^+}\ ,
\eea
which satisfy  the anticommutation relations

\be
\{\,{d^m}\,,\,{{\bar d}_n}\,\}\,=\,-i\,{\sqrt 2}\,{{\delta^m}_n}\,{\parp}\ .
\ee

To find an irreducible representation we have to impose the the chiral constraints

\be
{d^{\,m}}\,\phi\,=\,0\ ;\qquad {\bar d_{\,m}}\,\bar\phi\,=\,0\ ,
\ee
on a complex superfield $\phi(x^{\pm}, x, \bar x, \theta^m, \bar \theta_n)$. The solution  is then that 
\be
\phi = \phi({x^+}, y^-={x^-}-\,\frac{i}{\sqrt 2}\,{\theta_{}^m}\,{{\bar \theta}^{}_m}, x,\,{\bar x}, ~\theta^m).
\ee

We now have to add in $\theta$-terms into the Lorentz generators to complete the representation of the free algebra. The result is for $\lambda=0$

\be
j~=~x\,\bar\partial-\bar x\,\partial+ S^{12}_{}\ ,
\ee
where the little group helicity generator is 

\be
S^{12}_{}~=~ \,\frac{1}{ 2}\,(\,{\theta^p}\,{{\bar \partial}_p}\,-\,{{\bar \theta}_p}\,{\partial^p}\,)\,-\frac{i}{4\sqrt{2}\,\partial^+}\,(\,d^p\,\bar d_p-\bar d_p\,d^p\,).
\ee
 It ensures that the chirality constraints are preserved

\be
[\,j\,,\,d^m_{}\,]~=~[\,j\,,\,\bar d^{}_m\,]~=~0\ .
\ee
 
The other kinematical generators are 

\be
j^+_{}~=~i\, x\,\partial^+_{}\ ,\qquad \bar j^+_{}~=~i\,\bar x\,\partial^+_{}\ .
\ee
The rest of the generators must be specified separately for chiral and antichiral fields. Acting on $\phi$, we have 

\be
 j^{+-}_{}~=~i\,x^-_{}\,\partial^+_{}-\frac{i}{2}\,(\,\theta^p_{}\bar\partial^{}_p+\bar\theta^{}_p\,\partial^p_{}\,)\ ,
\ee
chosen so as to  preserve the chiral combination
\be
[\,j^{+-}_{}\,,\,y^-_{}\,]~=~-i\,y^-_{}\ ,
\ee
and such that its commutators  with the chiral derivatives

\be
[\,j^{+-}_{}\,,\,d^m_{}\,]~=~\frac{i}{2}\,d^m_{}\ ,\qquad [\,j^{+-}_{}\,,\,\bar d_m^{}\,]~=~\frac{i}{2}\,\bar d^{}_m\ ,
\ee
 preserve chirality. Similarly the dynamical boosts are 

\bea
j^-_{}&=&i\,x\,\frac{\partial\bar\partial}{\partial^+_{}} ~-~i\,x^-_{}\,\partial~+~i\,\Big( \theta^p_{}\bar\partial^{}_p\,+\frac{i}{4\sqrt{2}\,\partial^+}\,(\,d^p\,\bar d_p-\bar d_p\,d^p\,)\Big)\frac{\partial}{\partial^+_{}}\,\ ,\cr 
\bar j^-_{}&=&i\,\bar x\,\frac{\partial\bar\partial}{\partial^+_{}}~ -~i\,x^-_{}\,\bar\partial~+~ i\,\Big(\bar\theta_p^{}\partial_{}^p+\frac{i}{4\sqrt{2}\,\partial^+}\,(\,d^p\,\bar d_p-\bar d_p\,d^p\,)\,\Big)\frac{\bar\partial}{\partial^+_{}}\,\ .
\eea
They do not commute with the chiral derivatives, 

\be
[\,j^{-}_{}\,,\,d^m_{}\,]~=~\frac{i}{2}\,d^m_{}\,\frac{\partial}{\partial^+_{}}\ ,\qquad [\,j^{-}_{}\,,\,\bar d_m^{}\,]~=~\frac{i}{2}\,\bar d_m^{}\,\frac{\partial}{\partial^+_{}}\ ,
\ee
but do not change  the chirality of the fields on which they act.  They satisfy the Poincar\'e algebra, in particular

\be
[\,j_{}^-\,,\,\bar j^+_{}\,]~=~-i\,j^{+-}_{}-j\ ,\qquad [\,j^-_{}\,,\,j^{+-}_{}\,]~=~i\,j^{-}_{}\ .
\ee

We can now follow the same path as we did in the last section to go over to a "second-quantized" version in terms of integrals over the superfield and then add interaction terms to the dynamical generators and try to close the algebra. In this way we can construct all the known supersymmetric field theories as different representations of various supersymmetry algebras with different values of $\lambda$ and $N$. It is particularly interesting to study the cases $N=4 \times integer$. For those values one can impose a further condition on the superfield $\phi$ namely the "inside out" condition

\bea
\lefteqn{\bar d_{m_1}^{}\,\bar d_{m_2}^{}\,\,..\bar d_{m_{N/2-1}}^{}\,\bar d_{m_{N/2}}^{}\,\phi~=} 
 \nonumber \\  
&&\frac{1}{ 2} \epsilon_{{m_1}{m2}\,\,... {m_{N/2}} \,\,...{m_{N-1}}{m_N}^{}}\,d^{m_{N/2+1}}_{}\,d^{m_{N/2+2}}_{}\,\,...d^{m_{N-1}}_{}\,d^{m_N}_{}\,\bar\phi\ .
\eea

We can now construct three-point interaction terms for any $\frac{N}{4}$ even in the dynamical generators. This is certainly a tedious exercise based on writing the most general terms in the interaction terms and then check the full algebra. The resulting action is~\cite{Bengtsson:1983pg}

\bea 
S &= &\int d^4xd^N\theta d^N\bar \theta    \left\{\bar\phi(x,\theta) \frac{\Box}{{\partial^+}^{\frac{N}{2} }}\phi(x,\theta)  \right. \nonumber\\
& +& \frac{4g}{3}\Bigl[ \sum_{n=0}^\frac{N}{4} (-1)^n {\frac{N}{4} \choose n} \frac{1}{{\partial^+}^{N/2}}\bar\phi(x,\theta) {\bar \partial}^{\frac{N}{4} - n}{\partial^+}^n \phi(x, \theta)   \bar \partial^{n}{\partial^+}^{\frac{N}{4} -n}  \phi(x,\theta) \nonumber\\  
 &+&  c.c. \Bigr] \Bigr\} 
+ O(g^2).
\eea

When $\frac{N}{4}$ is odd, again the superfield has to transform as the adjoint representation of an external group with structure constants $f^{abc}$. The corresponding action is then

\bea 
S &= &\int d^4xd^N\theta d^N\bar \theta   \left\{\bar\phi^a(x,\theta) \frac{\Box}{{\partial^+}^{\frac{N}{2}} }\phi^a(x,\theta)\right.  \nonumber\\
& +& \frac{4g}{3} f^{abc}\Bigl[ \sum_{n=0}^\frac{N}{4} (-1)^n {\frac{N}{4} \choose n} \frac{1}{{\partial^+}^{N/2}}\bar\phi^a(x,\theta) {\bar \partial}^{\frac{N}{4} - n}{\partial^+}^n \phi^b(x, \theta)   \bar \partial^{n}{\partial^+}^{\frac{N}{4} -n}  \phi^c(x,\theta)  \nonumber\\  &+& c.c.\Bigr ] \Bigr\} 
+ O(g^2).
\eea

 We note that we can construct theories with higher spin if $\frac{N}{4}>2$. These are then very special combinations of the theories constructed in the previous section, with better quantum properties, since we know by experience that the more supersymmetry there is the better are the quantum properties.

\subsection{Maximally supersymmetric Yang-Mills Theory}

The case $N=4$ is especially interesting~\cite{Brink:1976bc}.  All the physical degrees of freedom are present in the superfield which can be expanded as  
\bea
\phi\,(y)&=&\frac{1}{ \partial^+}\,A\,(y)\,+\,\frac{i}{\sqrt 2}\,{\theta_{}^m}\,{\theta_{}^n}\,{\nbar C^{}_{mn}}\,(y)\,+\,\frac{1}{12}\,{\theta_{}^m}\,{\theta_{}^n}\,{\theta_{}^p}\,{\theta_{}^q}\,{\epsilon_{mnpq}}\,{\partial^+}\,{\bar A}\,(y)\cr
& &~~~ +~\frac{i}{\partial^+}\,\theta^m_{}\,\bar\chi^{}_m(y)+\frac{\sqrt 2}{6}\theta^m_{}\,\theta^n_{}\,\theta^p_{}\,\epsilon^{}_{mnpq}\,\chi^q_{}(y) \ .
\eea
The fields $A$ and $\bar A$ constitute the two helicities of a vector field while the antisymmetric $SU(4)$ bi-spinors $C^{}_{mn}$ represent six scalar fields since they satisfy

\bea
{{\nbar C}^{}_{mn}}~=~\,\frac{1}{2}\,{\epsilon^{}_{mnpq}}\,{C_{}^{pq}} \ .
\eea
The fermion fields are denoted by $\chi^m$ and $\bar\chi_m$. All have adjoint indices (not shown here), and are local fields in the  modified light-cone coordinates . This is the maximal supersymmetric Yang-Mills theory.  The full action is known~\cite{Brink:1982pd}

\bea
{\cal S}&=&-\int d^4x\int d^4\theta\,d^4 \bar \theta\,\Bigl\{\bar\phi^a\,\frac{\Box}{\partial^{+2}}\,\phi^a
+\frac{4g}{3}\,f^{abc}_{}\,\Big(\frac{1}{\partial^+_{}}\,\bar\phi^a_{}\,\phi^b_{}\,\bar\partial\,\phi^c_{}+{\rm c.c.}\Big)\cr
&&-g^2f^{abc}_{}\,f^{ade}_{}\Big(\,\frac{1}{\partial^+_{}}(\phi^b\,\partial^+\phi^c)\frac{1}{\partial^+_{}}\,(\bar \phi^d_{}\,\partial^+_{}\,\bar\phi^e)+\frac{1}{2}\,\phi^b_{}\bar\phi^c\,\phi^d_{}\,\bar\phi^e\Big)\, \Bigl\}.
\eea

With this action it was shown~\cite{Brink:1982wv} that the perturbation expansion is finite. There is no need for renormalization and the theory is very special. It is one of the cornerstones of modern particle physics. From the point of this lecture it appears as a very special representation of the superPoincar\'e algebra.

\subsection{Maximal Supergravity}

The next case is $N=8$~\cite{Cremmer:1978ds}. In this case the superfield can be expanded as

\bea
\phi\,(y)\,&=&\,\frac{1}{{\parp}^2}\,h\,(y)\,+\,i\,{\theta^{m}}\,{\frac {1}{{\parp}^2}}\,{{\bar \psi}_{m}}\,(y)\,+\,i\,{\theta^{m\,n}}\,{\frac {1}{\parp}}\,{{\bar A}_{m\,n}}\,(y)\nonumber \\
&&-\,{\theta^{m\,n\,p}}\,{\frac {1}{\parp}}\,{{\bar \chi}_{m\,n\,p}}\,(y)-\,{\theta^{m\,n\,p\,r}}\,{C_{m\,n\,p\,r}}\,(y)\,+\,i\,{\tilde \theta}^{(5)}_{m\,n\,p}\,{\chi^{m\,n\,p}}(y) \nonumber \\
&&+\,i\,{\tilde \theta}^{(6)}_{m\,n}\,\parp\,A^{m\,n}(y)+\,{\tilde \theta}^{(7)}_{m}\,\parp\,\chi^{m}(y)\,+\,{\tilde \theta}^{(8)}\,{\parp}^2\,{\bar h}\,(y)\ ,
\eea
where
\bea
\begin{split}
\theta^{{m_1}\,\ldots\,{m_n}}\,\equiv\,\frac{1}{n!}\,{\theta^{m_1}}\,\ldots\,{\theta^{m_n}},\ \;\;
{\tilde {\theta}}^{(n)}_{{n_1}\,\ldots\,{n_{8-n}}}\,\equiv\,\frac{1}{n!}\,\theta^{{m_1}\,\ldots\,{m_n}}\,\epsilon_{m_1\,\ldots\,m_n\,n_1\,\ldots\,n_{8-n}}\ .
\end{split}
\eea

The helicity in the field goes from $2$ to $-2$ and the theory has a spectrum comprised of a metric, twenty-eight vector fields, seventy scalar fields, fifty-six spin one-half fields and eight spin three-half fields. This theory is the maximal supergravity theory in $d=4$. The action can be simplified~\cite{Ananth:2005vg} to

\be 
S = \int d^4xd^8\theta d^8\bar \theta    \left\{\bar\phi(x,\theta) \frac{\Box}{{\partial^+}^4} \phi(x,\theta)  \right. 
 + \-\;\frac{3}{2}\,g\;{\frac {1}{{\parp}^2}}\;{\nbar \phi}\;\;{\bar \partial}\,{\phi}\;{\bar \partial}\,{\phi}+ 
 c.c. \Bigr] \Bigr\} 
+ O(g^2).
\ee

The computation of the four-point coupling is in progress~\cite{sud}. It is remarkable that the actions for the maximally supersymmetric Yang-Mills Theory and Supergravity Theory are so similar. In some sense the Supergravity Theory is just an extension of the Yang-Mills one. In the modern particle physics these two theories are very intimately connected even though the direct physical consequences of them look quite different.

\section{Light-Frame Formulations of Higher Dimensional Theories}

The procedure to find representations of the Poincar\'e algebra that we have followed in the previous section can, of course, be extended to field theories in dimensions of space-time higher than four. The covariant subalgebra which will be linearly realized is then $SO(d-2)$, so the physical fields will be representations of this algebra and hence characterized by these representations like we used helicity to distinguish the physical fields in four dimensions. If we just implement Poincar\'e invariance as in sect.2 we can, in principle, find all the possible field theories. However, the procedure gets easily tedious and furthermore there are few interesting quantum field theories in higher dimensions because of the renormalization problems. The only ones that are discussed are supersymmetric field theories since they are connected to the Superstring Theory. The ones that we have been interested in are the ones which lead to interesting field theories when compactified to four dimensions, so let us concentrate on those. The ones I will discuss here are ten-dimensional SuperYang-Mills and eleven-dimensional Supergravity which under compactification leads to the maximal theories discussed above.

\subsection{Ten-Dimensional SuperYang-Mill Theory}

The physical degrees of freedom of this theory are $\bf 8_v$ and an $\bf 8_s$. If we insist that the superfield should be a representation of the transverse $SO(8)$ it must be in one of the representations above. Since the natural spinor coordinate will also be an  $\bf 8_s$, such a superfield must include $ 8\times2^8$ components and must hence be very strongly restricted. Such a formalism has been developed~\cite{Brink:1983pf}, but it is not clear that the formalism is useful. Also it is not easily generalizable to the eleven-dimensional case. Instead I will describe a recent procedure developed in~\cite{Ananth:2004es}.

The idea is to use the same superfield as in four dimensions. In order to do that we have to sacrifice the explicit covariance under $SO(8)$ and use the decomposition

\be
SO(8)\supset~SO(2)~\times~SO(6)\ .
\ee

Since  $SO(6)\sim SU(4)$ we can identify the $SU(4)$ as the external symmetry group in the superfield equ. (46). The remaining  symmetry $SO(8)/(SO(6)\times SO(2))$ will transform among the components of the superfield. First of all,  the  transverse light-cone  space variables need be generalized to eight. We stick to the representions used in  the superfield, and introduce  the six extra coordinates  and their derivatives as antisymmetric bi-spinors

\be
{x^{m\,4}}\,=\,\frac{1}{\sqrt 2}\,(\,{x_{m\,+\,3}}\,+\,i\,{x_{m\,+\,6}}\,)\ ,\qquad 
{\partial^{m\,4}}\,=\,\frac{1}{\sqrt 2}\,(\,{\partial_{m\,+\,3}}\,+\,i\,{\partial_{m\,+\,6}}\,)\ ,
\ee
for $m\ne 4$, and their complex conjugates 

\be{{\bar x}_{pq}}\,=\,\frac{1}{2}\,{\epsilon_{pqmn}}\,{x^{mn}}\ ;\qquad{{\bar \partial}_{pq}}\,=\,\frac{1}{2}\,{\epsilon_{pqmn}}\,{\partial^{mn}}\ .
\ee
Their derivatives satisfy

\be
{{\bar \partial}_{mn}}\,{x^{pq}}\,~=~\,(\,{{\delta_m}^p}\,{{\delta_n}^q}\,-\,{{\delta_m}^q}\,{{\delta_n}^p}\,) \ ;\qquad 
{\partial^{mn}}\,{{\bar x}_{pq}}~=~(\,{{\delta^m}_p}\,{{\delta^n}_q}\,-\,{{\delta^m}_q}\,{{\delta^n}_p}\,) \ ,
\ee
and

\be
{{\partial}^{mn}}\,{x^{pq}}~=~\frac{1}{2}\,{\epsilon^{pqrs}}\,{{\partial}^{mn}}\,{{\bar x}_{rs}}~=~{\epsilon^{mnpq}}\ .
\ee
There are then no modifications to be made to the chiral superfield, except for the  dependence  on the extra coordinates

\be
A(y)~=~A(x,\bar x,x^{mn}_{},\bar x_{mn}^{},y^-_{})\ ,~~etc... ~\ .
\ee 
These extra variables  will be acted on by new operators that generate the higher-dimensional symmetries.

\subsection{The SuperPoincar\'e Algebra in $10$ Dimensions}
The SuperPoincar\'e algebra needs to be generalized from the  form in four dimensions.  One starts with the construction of the $SO(8)$ little group  using   the decomposition $SO(8)\supset SO(2)\times SO(6)$. The $SO(2)$ generator is the same; the $SO(6)\sim SU(4)$ generators are given by  

\bea
{j^m}_n\,&=&\,\frac{1}{ 2}\,(\,{x^{mp}}\,{{\bar \partial}_{pn}}\,-\,{{\bar x}_{pn}}\,{\partial^{mp}}\,)\,-\,{\theta^m}\,{{\bar \partial}_n}\,+\,{{\bar \theta}_n}\,{\partial^m}\,+\,\frac{1}{4}\,(\,{\theta^p}\,{{\bar \partial}_p}\,-\,{{\bar \theta}_p}\,{\partial^p}\,)\,{{\delta^m}_n}\cr
& &+ \frac{i}{2\sqrt{2}\,\partial^+}\,(\,d^m\,\bar d_n-\bar d_n\,d^m\,)+\frac{i}{8\sqrt{2}\,\partial^+}\,(\,d^p\,\bar d_p-\bar d_p\,d^p\,)\,{\delta^m}_{~n}\ .
\eea
 Note that we use the same spinors as in $4$ dimensions because of the  decomposition $SO(8)\supset SO(2)\times SO(6)$, where $SO(6)\sim SU(4)$.
The extra terms with the $d$ and $\bar d$ operators are not necessary for closure of the algebra. However they insure that the generators commute with the chiral derivatives. They satisfy the commutation relations

\bea
\Big[\,j\,,\,{j^m}_n\,\Big]~=~0\ ,\qquad 
\Big[\,{j^m}_n\,,\,{j^p}_q\,\Big]~=~{\delta^m}_q\,{j^p}_n-{\delta^p}_n\,{j^m}_q\ .
\eea
The remaining $SO(8)$ generators lie in the  coset  $SO(8)/(SO(2)\times SO(6))$ 

\bea
j^{pq}&=&x\,\partial^{pq}-x^{pq}\,\partial+\frac{i}{\sqrt 2}\,{\parp}\,{\theta^p}\,{\theta^q}-i\,{\sqrt 2}\,\frac{1}{\parp}\,{\partial^p}\,{\partial^q} +\frac{i}{\sqrt{2}\,\partial^+}\,d^p\,d^q\ ,\cr\cr
{{\bar j}_{mn}}&=&{\bar x}\,{{\bar \partial}_{mn}}-{{\bar x}_{mn}}\,{\bar \partial}+\frac{i}{{\sqrt 2}}\,{\parp}\,{{\bar \theta}_m}\,{{\bar \theta}_n}-i\,{\sqrt 2}\,\frac{1}{\parp}\,{{\bar \partial}_m}\,{{\bar \partial}_n}+\frac{i}{\sqrt{2}\,\partial^+}\,\bar d_m\,\bar d_n\ .
\eea
All   $SO(8)$ transformations are specially constructed so as not to mix chiral and antichiral superfields, 

\be
\,[\,{j^{mn}}\,,\,{{\bar d}_p}\,]\,~=~0\ ;\qquad \,[\,{{\bar j}_{mn}}\,,\,{d^p}\,]\,~=~0\ ,
\ee
and satisfy the $SO(8)$ commutation relations 

\beas
\Big[\,j\,,\,j^{mn}_{}\,\Big]&=&j^{mn}_{}\ ,\qquad \Big[\,j\,,\,\bar j_{mn}^{}\,\Big]~=~-\bar j_{mn}^{}\ ,\cr 
&&\cr
\Big[\,{j^m}_n\,,\,j^{pq}_{}\,\Big]&=&{\delta^q}_n\,j^{mp}_{}-\,{\delta^p}_n\,j^{mq}_{}\ ,\qquad
\Big[\,{j^m}_n\,,\,\bar j^{}_{pq}\,\Big]~=~{\delta^m}_q\,\bar j_{np}^{}\,-\,{\delta^m}_p\,\bar j_{nq}^{}\ ,\cr
&&\cr
{\Big[}\,{j^{mn}}\,,\,{{\bar j}_{pq}}\,{\Big ]}&=&{{\delta^m}_p}{{j^n}_q}\,+\,\,{{\delta^n}_q}{{j^m}_p}\,-\,{{\delta^n}_p}{{j^m}_q}\,-\,\,{{\delta^m}_q}{{j^n}_p}\,-\,(\,{{\delta^m}_p}\,{{\delta^n}_q}\,-\,{{\delta^n}_p}\,{{\delta^m}_q}\,)\,j\ .
\eeas
Rotations between the $1$ or $2$ and $4$ through $9$ directions induce on the chiral fields the changes

\be
\delta\,\phi~=~\Big(\,\frac{1}{2}\,\omega^{}_{mn}\,j^{mn}_{}+ \frac{1}{2}\,\bar\omega_{}^{mn}\,\bar j_{mn}^{}\,\Big)\,\phi\ ,
\ee
where complex conjugation is like duality

\be
{\bar \omega}_{pq}\,=\,\frac{1}{2}\,\epsilon^{}_{mnpq}\,\omega_{}^{mn}\ .
\ee
For example, a  rotation in the $1-4$ plane through an angle $\theta$ corresponds to taking $\theta=\omega_{14}=\omega_{23}$ ($ =\omega^{23}=\omega^{14}$ by reality), all other components being zero. Finally, we verify that the kinematical supersymmetries are duly rotated by these generators

\be
\,[\,{j^{mn}}\,,\,\bar q_{+\,p}^{}\,]~=~{{\delta^n}_p}\,{q^m_+}\,-\,{\delta^m}_p\,q^n_+ \ ;\qquad 
[\,{\bar j}_{mn}\,,\,{q^p_+}\,]~=~{{\delta_n}^p}\,{\bar q}_{+\,m}^{}\,-\,{\delta_m}^p\,{\bar q}_{+\,n}^{}\ .
\ee
We  now use the $SO(8)$ generators to construct the SuperPoincar\'e generators

\bea
j^+_{}&=&i\,x\,\partial^+_{}\ ;\qquad \bar j^{+}_{}~=~i\,\bar x\,\partial^+_{} \cr &&\cr
j^{+\,mn}_{}&=&i\,x^{mn}_{}\,\partial^+_{}\ ; \qquad 
\bar j^+_{~~mn}~=~i\,\bar x^{}_{mn}\,\partial^+_{}\ . 
\eea
The dynamical boosts are now 

\bea
j^-_{}&=&i\,x\,\frac{\partial\bar\partial\,+\,{\frac {1}{4}}\,{{\bar \partial}_{pq}}\,{\partial^{pq}}}{\partial^+_{}} ~-~i\,x^-_{}\,\partial+i\,{\frac { \partial}{\parp}}\,\Big\{\,{ \theta}^m\,{\bar\partial_m}~+~\frac{i}{4\sqrt{2}\,\parp}\,(d^p\,\bar d_p-\bar d_p\,d^p)\,\Big\}-\cr
&&~~~~~~~~~~~~~-~  {\frac {1}{4}}\,{\frac {\bar\partial_{pq}}{\parp}}\,{\biggl \{}\,\frac{\parp}{{\sqrt 2}}\,{{ \theta}^p}\,{{ \theta}^q}\,-\,\,\frac{\sqrt 2}{\parp}\,{{\partial}^p}\,{{ \partial}^q}+\frac{1}{\sqrt{2}\parp}d^p\,d^q\,\,{\biggr \}}\ ,
\eea
and its conjugate

\bea
\bar j^-_{}&=& i\,\bar x\,\frac{\partial\bar\partial\,+\,{\frac {1}{4}}\,{{\bar \partial}_{pq}}\,{\partial^{pq}}}{\partial^+_{}}-~i\,x^-_{}\,\bar\partial~+~i\,{\frac {\bar \partial}{\parp}}\,\Big\{\,{\bar \theta}_m\,{\partial^m}~+~\frac{i}{4\sqrt{2}\,\parp}\,(d^p\,\bar d_p-\bar d_p\,d^p)\,\Big\}-~\cr
 &&~~~~~~~~~~~~~-~ {\frac {1}{4}}\,{\frac {\partial^{pq}}{\parp}}\,{\biggl \{}\,\frac{\parp}{{\sqrt 2}}\,{{\bar \theta}_p}\,{{\bar \theta}_q}\,-\,\,\frac{\sqrt 2}{\parp}\,{{\bar \partial}_p}\,{{\bar \partial}_q}+\frac{1}{\sqrt{2}\parp}\bar d_p\,\bar d_q\,{\biggr \}}\ .
\eea
The others are obtained by using the $SO(8)/(SO(2)\times SO(6))$  rotations

\be
j^{-\,mn}_{}~=~[\,j^-_{}\,,\,j^{mn}_{}\,]\ ;\qquad 
\bar j^-_{~~mn}~=~[\,\bar j^-_{}\,,\,\bar j^{}_{mn}\,]\ .
\ee
We do not show their explicit forms as they are too cumbersome. 
The four supersymmetries in four dimensions turn into one supersymmetry in ten dimensions. In our notation, the kinematical supersymmetries  $q^n_+$ and $\bar q^{}_{+n}$, are assembled into one $SO(8)$ spinor. The dynamical supersymmetries are obtained by boosting 

\be
i\,[\,{\bar j}^-\,,\,q_+^m\,]~\equiv~{\cal Q}^m_{}\ ,\qquad 
i\,[\,j^-\,,\,\bar q_{+\,m}\,]~\equiv~\nbar{\cal Q}_m^{}\ ,
\ee
where 

\bea
{\cal Q}^m_{}&=&{\frac {\bar \partial}{\parp}}\,{{q_+}^m}\,+\,{\frac {1}{2}}\,{\frac {\partial^{mn}}{\parp}}\,{{\bar q}_{+\,n}}\ ,\cr
&&\cr
\nbar{\cal Q}_m^{}&=&\frac {\partial}{\parp}}\,{{\nbar q}_{+\,m}}\,+\,{\frac{1}{2}}\,{\frac {{\nbar \partial}_{mn}}{\parp}\, q^{~n}_{\,+}\ .
\eea
They satisfy the supersymmmetry algebra

\be
\{\,{\cal Q}^{\,m}_{}\,,\,\nbar {\cal Q}^{}_{\,n}\,\}~=~i\,\sqrt{2}\,\delta^m_{~n}\,\frac{1}{\parp}
\,\Big(\partial\,{\nbar \partial}\,+\,\frac{1}{4}\,{\nbar \partial}_{pq}\,\partial^{pq}\,\Big)\ ,
\ee
and can be  obtained from one another by $SO(8)$ rotations, as 
\be
\frac{1}{2}\,{\epsilon_{pqmn}}\;[\,j^{pq}\,,\,{\cal Q}^{\,m}\,]\;=\;~4\,{\nbar {\cal Q}}_{\,n}\ ,
\ee
while 
\be
[\,\bar j_{pq}\,,\,{\cal Q}^{\,m}\,]~=~0\ .
\ee

Note also that 

\be
\{\,{\cal Q}^{\,m}_{}\,,\,q_+^n\,\}~=~\frac{i}{\sqrt 2}\, \partial^{mn}\ ,
\ee
 
 \subsection{The Generalized Derivatives}
The cubic interaction in the  $N=4$ Lagrangian contains explicitly the  derivative operators $\partial$ and $\bar\partial$. To achieve covariance in ten dimensions, these must be generalized. We propose the following operator 
\be
{\nbar \nabla}~\equiv~{\bar \partial}\,+\,\frac{i\,\alpha}{4\,\sqrt 2\,\partial^+}\,{{\bar d}_p}\,{{\bar d}_q}\,\partial^{pq} \ ,
\ee
which naturally incorporates the rest of the derivatives $\partial^{pq}$, with   $\alpha$ as an arbitrary parameter. After some algebra, we find that $\nbar\nabla$ is covariant under $SO(8)$ transformations. We define its rotated partner as 

\be
{\nabla_{}^{mn}}~\equiv~{\Big[}\,{\nbar \nabla}\,,\,{j^{mn}}\,{\Big]}\ , 
\ee
where

\be
{\nabla_{}^{mn}}~=~{\partial^{mn}_{}}\, -\,\frac{i\,\alpha}{4\,\sqrt 2\,\parp}\,{{\bar d}^{}_r}\,{{\bar d}^{}_s}\,{\epsilon_{}^{mnrs}}\,\partial\ .
\ee
If we apply to it the inverse transformation, it goes back to the original form 

\be
\Big [\,{\nbar j}_{pq}\,,\,\nabla^{mn}\,\Big] \,~=~\,(\,{\delta_p}^m\,{\delta_q}^n\,-{\delta_q}^m\,{\delta_p}^n\,)\,\nbar \nabla\ ,
\ee
and these operators transform under   $SO(8)/(SO(2)\times SO(6))$, and $SO(2)\times SO(6)$  as  the components of an $8$-vector.  

We introduce the conjugate operator $\nabla$ by requiring that 

\be
\nabla\,\bar\phi~\equiv~ \nbar{(\nbar\nabla\,\phi)}\ ,
\ee
with  

\be
{ \nabla}~\equiv~{\partial}\,+\,\frac{i\,\alpha}{4\,\sqrt 2\,\partial^+}\,{{ d}^p}\,{{ d}^q}\,\bar\partial_{pq} \ .
\ee
Define 

\be
{\nbar \nabla}_{\;mn}~{\equiv}~{\Big[}\,{{\nabla}}\,,\,{{\bar j}_{mn}}\,{\Big]}\ ,
\ee
which is given by 
\be
{\nbar \nabla^{}_{mn}}~=~{\bar\partial_{mn}^{}}\, -\,\frac{i\,\alpha}{4\,\sqrt 2\,\parp}\,{{ d}^r_{}}\,{{ d}_{}^s}\,{\epsilon_{mnrs}^{}}\,\bar \partial\ .
\ee
We then  verify that 

\be
\Big [\,{ j}^{mn}\,,\,\nbar\nabla_{pq}\,\Big] \,~=~\,(\,{\delta_p}^m\,{\delta_q}^n\,-{\delta_q}^m\,{\delta_p}^n\,)\, \nabla\ .  
\ee
 
The kinetic term is trivially made $SO(8)$-invariant by including the six extra transverse derivatives in the d'Alembertian. The quartic interactions are  obviously invariant since they  do not contain any transverse derivative operators. Hence we need only consider the cubic vertex. 
In the paper~\cite{Ananth:2004es} it is shown that to achieve covariance in ten dimensions, it suffices indeed to replace the transverse $\partial$ and $\bar\partial$ by $\nabla$ and $\nbar\nabla$, respectively.  This is done by checking the invariance under the little group $SO(8)$. Together with the result from four dimensions this is enough to warrant invariance under the full superPoincar\'e group in ten dimensions. In this process the parameter $\alpha$ is determined to be $1$. The full  action is then

\bea
{\cal S}&=&-\int d^4x\int d^4\theta\,d^4 \bar \theta\,\Bigl\{\bar\phi^a\,\frac{\Box}{\partial^{+2}}\,\phi^a
+\frac{4g}{3}\,f^{abc}_{}\,\Big(\frac{1}{\partial^+_{}}\,\bar\phi^a_{}\,\phi^b_{}\,\bar\nabla\,\phi^c_{}+{\rm c.c.}\Big)\cr
&&-g^2f^{abc}_{}\,f^{ade}_{}\Big(\,\frac{1}{\partial^+_{}}(\phi^b\,\partial^+\phi^c)\frac{1}{\partial^+_{}}\,(\bar \phi^d_{}\,\partial^+_{}\,\bar\phi^e)+\frac{1}{2}\,\phi^b_{}\bar\phi^c\,\phi^d_{}\,\bar\phi^e\Big)\, \Bigl\}.
\eea

This action is suitable in order to investigate the perturbative properties of the theory. It is, of course, non-renormalizable but has still remarkable properties that Nature might use. One can also study possible higher symmetries of this action.
 
\subsection{Eleven-Dimensional Supergravity}
 
$N=1$ Supergravity in eleven dimensions, contains three different massless fields, two bosonic (gravity and a three-form) and one Rarita-Schwinger spinor. Its physical degrees of freedom are classified in terms of the transverse little group, $SO(9)$, with the graviton $G^{(\,MN\,)}$, transforming as a symmetric second-rank tensor, the three-form $B^{[\,MNP\,]}$ as an anti-symmetric third-rank tensor and the Rarita-Schwinger field as a spinor-vector, $\Psi^M$ ($M,N,\ldots$ are $SO(9)$ indices). This theory on reduction to four dimensions leads to the maximally supersymmetric $N=8$ theory.

In order to use the formalism and especially the superfield equ. (49) developed in four dimensions for the maximally supersymmetric $N=8$ theory we have to decompose 

\be
SO(9)\supset~SO(2)~\times~SO(7)\ .
\ee
 
  The $SO(7)$ symmetry  can in fact be upgraded to an $SU(8)$ symmetry. However, it is important to remember that it is really the $SO(7)$ which is relevant when we ``oxidize" the theory to $d\,=\,11$ and the coordinates $\theta^m$ and $\bar \theta_n$ used in the four-dimensional case will now be interpreted as spinors under $SO(7)\times SO(2)$. To distinguish this we will change the notation $m,\,n$ to $\alpha, \,\beta$ for the spinors and use the notation $a,\,b$ for the vector indices of $SO(7)$.

The first step is to generalize the transverse variables to nine. In the Yang-Mills case, the compactified $SO(6)$ was easily described by $SU(4)$ parameters and we made use of the convenient bi-spinor notation. In the present case, the compactified $SO(7)$ has no equivalent unitary group so we simply introduce additional real coordinates, ${x^a}$ and their derivatives $\partial^a$(where $a$ runs from $4$ through $10$). The chiral superfield remains unaltered, except for the added dependence on the extra coordinates
\be
h(y)~=~h(x,\bar x,{x^a},y^-_{})\ ,~~etc... ~\ .
\ee 
These extra variables  will be acted on by new operators that will restore the higher-dimensional symmetries.

\subsection{The SuperPoincar\'e Algebra in $11$ Dimensions}
The SuperPoincar\'e algebra needs to be generalized from its four-dimensional version. The $SO(2)$ generators stay the same and we propose generators of the coset $SO(9)/(SO(2)\times SO(7))$, of the form,
\bea
{j^a}\,&=&\,-\,i\,(\,x\,{\partial^a}\,-\,{x^a}\,{\partial}\,)\,+\,{\frac {i}{2\,\sqrt 2}}\,\parp\;{\theta^\alpha}\,{{(\,{\gamma^a})}_{\alpha\,\beta}}\,{\theta^\beta}\,-\,{\frac {i}{\sqrt 2\,\parp}}\;{\partial^\alpha}\,{{(\,{\gamma^a})}_{\alpha\,\beta}}\,{\partial^\beta}\, \nonumber \\
&&+\,{\frac {i}{2\,\sqrt 2\,{\parp}}}\;{d^\alpha}\,{{(\,{\gamma^a})}_{\alpha\,\beta}}\,{d^\beta}
\eea

\bea
{{\nbar j}^{\;b}}\,&=&\,-\,i\,(\,{\bar x}\,{\partial^b}\,-\,{x^b}\,{\bar \partial}\,)\,+\,{\frac {i}{2\,\sqrt 2}}\,\parp\;{{\bar \theta}_\alpha}\,{{(\,{\gamma^b})}^{\alpha\,\beta}}\,{{\bar \theta}_\beta}\,-\,{\frac {i}{\sqrt 2\,\parp}}\;{{\bar \partial}_\alpha}\,{{(\,{\gamma^b})}^{\alpha\,\beta}}\,{{\bar \partial}_\beta}\, \nonumber \\
&&+\,{\frac {i}{2\,\sqrt 2\,{\parp}}}\;{{\bar d}_\alpha}\,{{(\,{\gamma^b})}^{\alpha\,\beta}}\,{{\bar d}_\beta}
\eea
which satisfy the $SO(9)$ commutation relations,

\bea
\Big[\,j\,,\,j^a\,\Big]&=&j^a\ ,\qquad \Big[\,j\,,\,\bar j^b\,\Big]~=~-\bar j^b \nonumber \\
\Big[\,j^{cd}\,,\,j^a\,\Big]&=&\delta^{ca}\,j^d\,-\,\delta^{da}\,j^c \nonumber \\
{\Big[}\,j^a\,,\,\bar j^b\,{\Big ]}&=&\,i\,j^{ab}\,+\,\delta^{ab}\,j,
\eea
where $j$ is the same as before,  and the $SO(7)$ generators read,
\bea
j^{ab}\,&=&\,-\,i\,(\,x^a\,\partial^b\,-\,x^b\,\partial^b\,)\,+\,\theta^\alpha\,{(\gamma^a)}^{\alpha\,\,\beta}\,{(\gamma^b)}^{\beta\,\,\sigma}\,{\bar \partial}_\sigma\, \nonumber \\
&&+\,{\bar \theta}_\alpha\,{(\gamma^a)}^{\alpha\,\,\beta}\,{(\gamma^b)}^{\beta\,\,\sigma}\,{\partial}^\sigma\,-\frac{1}{\sqrt 2\,\parp}\,{d^\alpha}\,{(\gamma^a)}^{\alpha\,\,\beta}\,{(\gamma^b)}^{\beta\,\,\sigma}\,{\bar d}_\sigma\ .
\eea
The full $SO(9)$ transverse algebra is generated by $j\,,\,j^{ab}\,,\,{j^a}$ and ${\bar j}^b$. All rotations are specially constructed to preserve chirality. For example,
\be
\,[\,{j^{a}}\,,\,{{\bar d}_{\alpha}}\,]\,~=~0\ ;\qquad \,[\,{{\bar j}^{\,b}}\,,\,{d^\alpha}\,]\,~=~0\ .
\ee
The remaining kinematical generators do not get modified,
\bea
j^+\,=\,j^+\ ,\qquad j^{+-}\,=\,j^{+-}\ ,
\eea
while new kinematical generators appear,
\bea
j^{+\,a}&=&i\,x^{a}\,\partial^+_{}\ ; \qquad 
\bar j^{+\,b}~=~i\,\bar x^{b}\,\partial^+_{}\ . 
\eea
We generalize the linear part of the dynamical boosts to, 
\bea
j^-_{}&=&i\,x\,\frac{\partial\bar\partial\,+\,{\frac {1}{2}}\,{\partial^a}\,{\partial^a}}{\partial^+_{}} ~-~i\,x^-_{}\,\partial+i\,{\frac { \partial}{\parp}}\,\Big\{\,{ \theta}^\alpha\,{\bar\partial_\alpha}~+~\frac{i}{4\sqrt{2}\,\parp}\,(d^\alpha\,\bar d_\alpha-\bar d_\alpha\,d^\alpha)\,\Big\} \nonumber \\
&&-\,{\frac {1}{4}}\,{\frac {\partial^a}{\parp}}\,{\Big \{}\,{\parp}\;{\theta^\alpha}\,{{(\,{\gamma^a})}_{\alpha\,\beta}}\,{\theta^\beta}\,-\,{\frac {2}{\parp}}\,\;{\partial^\alpha}\,{{(\,{\gamma^a})}_{\alpha\,\beta}}\,{\partial^\beta}\,+\,{\frac {1}{\parp}}\,\;{d^\alpha}\,{{(\,{\gamma^a})}_{\alpha\,\beta}}\,{d^\beta}\,{\Big \}}. \nonumber \\
\,
\eea

The other boosts may be obtained by using the $SO(9)/(SO(2)\times SO(7))$ rotations,

\be
j^{-\,a}_{}~=~[\,j^-_{}\,,\,j^{a}_{}\,]\ ;\qquad 
\bar j^{-\,b}~=~[\,\bar j^-_{}\,,\,\bar j^{b}\,]\ .
\ee
We do not show their explicit forms as they are too cumbersome. The dynamical supersymmetries are obtained by boosting

\bea
\label{ds}
[\,j^-\,,\,\bar q_{+\,\eta}\,]~\equiv~\nbar{\cal Q}_\eta^{}&=&-\,i\,\frac {\partial}{\parp}}\,{{\nbar q}_{+\,\eta}}\,-\,{\frac{i}{\sqrt 2}}\,{{(\,{\gamma^b}\,)}_{\,\eta\,\rho}}\,q^{~\rho}_{\,+}\,{\frac {\partial^b}{\parp}\ ,\cr
&&\cr
[\,{\bar j}^-\,,\,q_+^\alpha\,]~\equiv~{\cal Q}^\alpha_{}&=&i\,{\frac {\bar \partial}{\parp}}\,{{q_+}^\alpha}\,+\,{\frac {i}{\sqrt 2}}\,{{(\,{\gamma^a}\,)}^{\,\alpha\,\beta}}\,{{\bar q}_{+\,\beta}^{}}\,{\frac {\partial^a}{\parp}}\ .
\eea

They satisfy, 
\bea
\{\,{\cal Q}^{\,\alpha}_{}\,,\,q_+^\eta\,\}~=~-\,{{(\,{\gamma^a}\,)}^{\alpha\,\eta}}\,{\partial^a}\ ,
\eea

and the supersymmetry algebra,

\bea
\{\,{\cal Q}^{\,\alpha}_{}\,,\,\nbar {\cal Q}^{}_{\,\eta}\,\}~=~i\,\sqrt{2}\,\;{{\delta^{\alpha}}_{\eta}}\,\frac{1}{\parp}
\,\Big(\partial\,{\nbar \partial}\,+\,\frac{1}{2}\,{\partial^a}\,{\partial^a}\,\Big)\ .
\eea

Having constructed the free $N=1$ SuperPoincar\'e generators in eleven dimensions which act on the chiral superfield, we turn to building the interacting theory.

\subsection{The Generalized Derivatives}
The cubic interaction in the $N=8$ Lagrangian explicitly contains the  transverse derivative operators $\partial$ and $\bar\partial$. To achieve covariance in eleven dimensions, we proceed to generalize these operators as we did for $N=4$  Yang-Mills. We propose the generalized derivative 

\bea
{\nbar \nabla}\;=\;{\bar \partial}\,+\,{\frac {\sigma}{16}}\,{{\bar d}_{\alpha}}\,{{(\,{\gamma^a}\,)}^{\alpha\,\beta}}\,{{\bar d}_\beta}\,{\frac {\partial^a}{\parp}}\,,
\eea
which naturally incorporates the coset derivatives $\partial^{m}$. Here $\sigma$ is a parameter, still to be determined. We use the coset generators to produce its rotated partner $\nbar\nabla$ by,

\bea
[\;{\nbar \nabla}\,,\,{j^a}\;]\;{\equiv}\;{\nabla^a}\;=\;-\,i\,{\partial^a}\,+\,{\frac {i\,\sigma}{16}}\,{{\bar d}_{\alpha}}\,{{(\,{\gamma^a}\,)}^{\alpha\,\beta}}\,{{\bar d}_\beta}\,{\frac {\partial}{\parp}}\,.
\eea

 It remains to verify that the original derivative operator is reproduced by undoing this rotation; indeed we find the required closure,

\beas
[\;{\nabla^a}\,,\,{{\nbar j}^{\;b}}\;]\;=\;{\delta^{\,a\,b}}\,\;{\nbar \nabla}
\eeas

\vskip 0.3cm

The new derivative $(\;{\nbar \nabla}\,,\,{\nabla^a}\;)$, thus transforms as a 9-vector under the little group in eleven dimensions. We note that $\sigma$ is not determined by these algebraic requirements. Instead, its value will be fixed by requiring that our generalized vertex satisfy the correct invariance requirements. We define the conjugate derivative $\nabla$, by requiring that 

\be
\nabla\,\bar\phi~\equiv~\nbar{(\nbar\nabla\,\phi)}\ .
\ee
This tells us that, 

\bea
\nabla~\equiv~\;{\partial}\,+\,{\frac {\sigma^*}{16}}\,{d^\alpha}\,{{(\,{\gamma^b}\,)}^{\alpha\,\beta}}\,{d^\beta}\,{\frac {\partial^b}{\parp}}\,
\eea 
This construction is akin to that for the $N=4$ Yang-Mills theory, but this time it applies to the ``oxidation" of the ($N=8$, $d=4$) theory to ($N=1$, $d=11$) Supergravity. This points to remarkable algebraic similarities between the two theories, with possibly profound physical consequences. It remains to show that the simple replacement of the transverse derivatives $\partial,\bar \partial$ by $\nabla,\nbar \nabla$ in the ($N=8$, $d=4$) interacting theory yields the fully covariant Lagrangian in eleven dimensions. 
 
This can be done by checking the invariance under the little group $SO(9)$. This is a very tedious exercise which was done in paper~\cite{Ananth:2005vg}. Indeed it is possible to show  that the three-point coupling is invariant for the specific choice of $\sigma=-\sqrt 2$ and the eleven-dimensional supergravity theory can be written as 
 
\be 
S = \int d^{10}x d^8\theta d^8\bar \theta  \left\{\bar\phi(x,\theta) \frac{\Box}{{\partial^+}^4} \phi(x,\theta)  \right. 
 + \-\;\frac{3}{2}\,g\;{\frac {1}{{\parp}^2}}\;{\nbar \phi}\;\;{\nbar \nabla}\,{\phi}\;{\nbar \nabla}\,{\phi}+ 
 c.c. \Bigr] \Bigr\} 
+ O(g^2).
\ee

The computation of the four-point coupling is in progress. With it one can study various properties of this theory, such as the one-loop graphs. They will diverge but there might be ways to add more fields to get convergent answer. This is the long term goal of this project. One can also study the symmetries of the action. It is clear that the action is quite unique and has a profound r\^ole in modern particle physics and any symmetry that can be found for this action is a genuine physical symmetry. This theory is also the low-energy limit of the mystic M-theory which is supposed to be the underlying theory to all string theories. This theory is shrouded in mystery and any attempt to better understand the supergravity theory  can help us eventually understand M-theory.

\section{Strings}

We have so far studied the physics of point-particles. We can ask what happens if we extend this programme to also study one-dimensional objects, strings. The fields will then be functionals $\Phi(x(\sigma))$ and the corresponding theories will be functional field theories, which is a subject much less understood. However, we can get far by studying the "first-quantized" version.

The relativistic dynamics of extended bodies is quite difficult to handle. There is the severe problem of simultaneity. A string that moves will in general have different times along it. There is one exception, though, and that is if we choose "time" to be one of the light-cone directions. We can, in fact, choose the same $x^+$ along the string. But this is just the time we like to work with! We also have to specify the boundary conditions for the string. Either we choose an open string or a closed string. We here treat them both but specify the length to be $\pi$. We introduce a momentum density $p^\mu$ and demand the commutation rules

\be
[x^\mu(\sigma),p^\nu(\sigma^\prime)]= i \delta(\sigma -\sigma^\prime) \eta^{\mu\nu}
\ee
 
We can now try to imitate the algebra in the point-particle case and try the following generators at $x^+=0$

\bea
p^+ &=& p^+, \\ \nonumber
p^- &=& \frac{1}{2p^+}\int^\pi_0 d\sigma \,\bigl({(p^i(\sigma)})^2 +({\xprim}^i(\sigma))^2\bigr), \\ \nonumber
p^i &=& \int^\pi_0 d\sigma \,p^i(\sigma), \\ \nonumber
j^{ij} &=& \int^\pi_0 d\sigma\, \bigl(x^i(\sigma) p^j(\sigma)- (x^j(\sigma) p^i(\sigma) \bigr), \\ \nonumber
j^{+j} &=& -x^i p^+,\\ \nonumber
j^{+-} &=& -x^- p^+,\\ \nonumber
j^{-i} &=&   \int^\pi_0 d\sigma\, \bigl(x^-(\sigma) p^i(\sigma)- (x^i(\sigma) p^-(\sigma) \bigr).
\eea

In this expression $i, \,j$ denote the transverse directions and $p^-(\sigma)$ is the integrand of $p^-$. The function $x^-(\sigma)$ is an unknown function, which will be determined such that the algebra closes. The closure of this algebra was established in the famous paper of Goddard, Goldstone, Rebbi and Thorn~\cite{Goddard:1973qh}. The remarkable result is that it only closes if the dimension of space-time is $d=26$. This was a very surprising result at the time.

Consider an open string with the parametrization 

\be
x^i(\sigma) = x^i + i \sum_{n\neq 0} \frac{1}{n} \alpha^i_n\, cos n\sigma,
\ee
 
 where the oscillator modes $\alpha$ satisfy
 
\be
[\alpha^i_m, \alpha^j_n]= m\delta_{m+n,0} \delta^{ij}.
\ee
 
Insert this into the generator $p^-$.

\be
p^-= \frac{{p^i}^2}{2p^+} + \frac{1}{p^+} \sum_{n=1}^\infty \alpha^i_n \alpha^i_n.
\ee
 
or 

\be
p^2= -m^2= -  \sum_{n=1}^\infty \alpha^i_n \alpha^i_n.
\ee

The string constitutes an infinite set of harmonic oscillators. If we compute the lowest mass state we will only have the zero-mode fluctuations of all the oscillators~\cite{Brink:1986ja}. The frequency is essentially $n$ and the computation gives

\be
m^2= \frac{d-2}{2} \sum_{n=1}^\infty n =  \frac{d-2}{2} \sum_{n=1}^\infty n^{-s} \mid_{s=-1} = - \frac{d-2}{24}.
\ee
In this calculation we used a $\zeta$-function renormalization of the infinite sum. In the orginal paper~\cite{Brink:1986ja} we used a renormalization of the velocity of light which is the velocity with which the phonons on the string travel.
 
The spin-$1$ state must be massless so this scalar state is a tachyon with   ${mass}^2=-1$. (We have suppressed a mass scale, which does not affect the argument). Again we see that it only works for $d=26$.

We note that a string formalism is much more constrained than the corresponding one for point-like particles. The formalism above is a representation of the Poincar\'e algebra but it is unphysical. 

How can one find a representation which is physical in the sense of not having any tachyons? Equ. (110) is the crucial one to understand. The only way is to cancel the negative contribution from the zero-point fluctuations with a corresponding positive contribution. We know that the zero-point flucutation from a fermionic oscillator is negative so an infinite sum like the one above over fermionic oscillators could cancel the term from the bosonic ones. Suppose we introduce a Grassmann coordinate $\lambda^{\mu}(\sigma)$. It should lead to a cancellation if $\lambda^{\mu}$ satisfy the same boundary conditions as $x^\mu$. In fact scrutinizing possible boundary conditions one finds indeed that such conditions are possible, but there is also another sector with modes with half-integer frequencies. This sector would again lead to tachyons. The two sectors found are the Ramond~\cite{Ramond:1971gb} and the Neveu-Schwarz~\cite{Neveu:1971rx} sectors. One could also try to use a spinorial coordinate $\theta^{\alpha}(\sigma)$ . This can only be done in $d=3\,,4\,,6$ or $10$, since in the transverse space the vector and the spinor then have the same dimension. The latter model is the Superstring Model~\cite{Green:1980zg}.

Adding in the Grassmann coordinates into the Poincar\'e generators one finds that the algebra does indeed close in $d=10$. Furthermore one finds that one can construct the full  $d=10$ superPoincar\'e algebra. By playing with open and closed strings and combining the  $d=10$ with the  $d=26$ model one finds that one can construct five different string theories. These have been the basis for much of the work in string theory for the last twenty years. 

It is remarkable that there exist only five different physically consistent representations of the superPoincar\'e algebra in terms of strings. They can all be found quite simply by just trying to construct representations of the Poincar\'e algebra. For a detailed study of this method, see~\cite{Brink:1985bb}.
 
\section{Continuous Spin Representations}

There are other representations found by Wigner that do not seem to be realized in Nature. These are the ``continuous spin representations" (CSR), which describe a massless object with an infinite number of helicities. Wigner himself argued~\cite{WIGNER2} against their use in physics since they lead to infinite heat capacity of the vacuum.  

They are found by scrutinizing the generators in sect.~2 writing them in an arbitrary dimension. In fact they are not the most general. In fact one can add an operator $T^i$ into $j^{-i}$, $i$ being the transverse index.

\be
j_{}^{-i}~=~x_{}^-p_{}^i-\frac{1}{2}\{x^i,p_{}^-\}+\frac{1}{p^+}
(T_{}^i-p_{}^js_{}^{ij})\ ,
\ee
where the $T^i_{}$   transform as $SO(d-2)$ vectors
\be
[\,s^{ij}_{}\,,\,T^k_{}\,]~=~i\delta^{ik}_{}T^j_{}-i\delta^{jk}_{}T^l_{}\
,
\ee
and satisfy
\be
[\,T^i_{}\,,\,T^j_{}\,]~=~im_{}^2s^{ij}_{}\ 
.\ee
When $m\ne 0$,  $T^i/m$ are the generators of $SO(d-1)/SO(d-2)$, which together with  $S^{ij}$, complete  the massive little group $SO(d-1)$. 
 When $m=0$, the $T^i$  commute with one another, acting  as light-cone
translations, and the algebra can be satisfied in two ways: 
\begin{itemize}
  \item $T^i=0$. This corresponds to the familiar massless representations
which describe particles with a finite number of degrees of freedom,
realized on states that  satisfy
\be
T^i\vert~p^+,p^i;\, (a_1,\dots,a_{r})\,>~=~0\ ,\ee
where $(a_1,\dots,a_r)$ are the Dynkin labels of  $SO(d-2)$
representations and $r$ is the rank of the little group. These label 
the different helicity states of the
massless particle. In four dimensions, the Pauli-Lubanski vector is
light-like.
\item $T^i\ne 0$. In this case, $T^i$ are the $c$-number components of a 
transverse vector. The states on which the Poincar\'e algebra 
is realized are
\be
T^i\vert~p^+,p^i;\, \xi^i,\, 
(a_1,\dots,a_{r})\,>~=~\xi^i~\vert~p^+,p^i;\,\xi^i,\,
(a_1,\dots,a_{r})\,>\ ,
\ee
which have additional labels, in the form of a little group vector
$\xi^i$. There is an important difference from the previous case, since $(a_1,\dots,a_{r})$ now labels  the $SO(d-3)$ subgroup of the 
transverse little group $SO(d-2)$. In four dimensions, there is no such group and the states are simply labelled by an additional space-like vector of constant magnitude.  
These  span two distinct representations,  called ``continuous 
spin representations" by Wigner in his original work {\cite{Wigner:1939cj}}. They are characterized by a space-like Pauli-Lubanski vector, and describe a massless state with an infinite number of integer-spaced helicities. 
\end{itemize}

It is interesting to ask what happens to these representations in dimensions higher than four and when supersymmetry is added in~\cite{Brink:2002zx}. Firstly the two representations above in $d=4$ span either integer helicities or half-integer helicities and the sum of the two is really the representation of the superPoincar\'e algebra. Unlike $(3+1)$ dimensions, there are infinite numbers of CSRs in higher dimensions. The states are as said above no longer characterized by the light-cone little group, but by its subgroup orthogonal to $T^i$, which we call the 'short little group', in order to have a commuting set of operators to classify the states. For arbitrary dimensions the representations are hence labeled both by the length of a space-like  translation vector (the eigenvalue of $T^i$) and the Dynkin indices of the short little group $SO(d-3)$.  Continuous spin representations are in one-to-one correspondence with representations of the short little group. 
 
 Let us consider specifically the case of eleven dimensions. It is particularly interesting because of its connection to the elusive M-theory.  
In eleven dimensions the short little group  is $SO(8)$ that leaves the light-cone translation 
vector  $\vec{T}$ invariant.
We decompose the supercharge in $d=11$ into two 8-component supercharge, $\cq^a_+$ 
and $\cq^{\dot{a}}_+$, each transforming as a different $SO(8)$ spinor.  They can both be used to build up supermultiplets that will be representations of the superPoincar\'e algebra. These are well known in $10$-dimensional superphysics and we would get the representations of type IIA and IIB supergravity which have $N=2$ supersymmetry as well as the $N=1$ representations of which the simplest is the SuperYang-Mills above. These representations will then all be increased to be Continuous Spin Representations.

So far these representations have not entered into any realistic physics model but they have appeared in an interesting study of tensionless strings by Savvidy~\cite{Savvidy:2003fx}.

\section{Concluding remarks}

In this lecture I have shown that all the known quantum field theories follow by studying representations of the Poincar\'e algebra. What we get though is essentially the part of them which is amenable to perturbation theory, ie as expansions in a coupling constant. We have learnt in recent years that quantum field theories are very much richer than what meets the eye in a perturbation expansion. The formalism here is not suitable for such studies. It is very hard if possible to study non-perturbative effects such as solitons, magnetic monopoles, branes and various forms of duality. However, the formalism is a complement to other studies, and it is very useful for certain studies about finiteness in perturbations expansions, which is one of the crucial tests of a quantum gravity theory.

\end{document}